\documentclass[10pt,twocolumn]{article}

\usepackage{amsmath,amssymb,bbm}
\usepackage{mathrsfs,enumerate,latexsym,graphicx}
\usepackage{hyperref}
\usepackage{fullpage}
\usepackage{abstract}
\usepackage[]{appendix}
\usepackage{doi}

\usepackage[numbers]{natbib}

\usepackage{blindtext}

\usepackage[T1]{fontenc} % need this to make the pipe symbol | for writing Wolfram|Alpha

\usepackage{authblk}
\makeatletter
\renewcommand\AB@affilsepx{ \protect\Affilfont}
\makeatother

\newtheorem{definition}{Definition}

\usepackage[usenames,dvipsnames]{xcolor} % extra colors http://en.wikibooks.org/wiki/LaTeX/Colors

\newcommand{\new}[1]{{{#1}}}
\newcommand{\newA}[1]{{{#1}}}
\newcommand{\newB}[1]{{{#1}}}
\newcommand{\newC}[1]{{{#1}}}

\newcommand{\newfinal}[1]{{{#1}}}

\newcommand{\up}{X}
\newcommand{\inter}{Y}
\newcommand{\down}{Z}

\newcommand{\drives}{\rightharpoonup}

\newcommand{\strength}{\sigma}

\newcommand{\coupling}[3]{\mathcal{C}_{#3}({#1}, {#2})}

\newcommand{\diff}{\mathrm{d}}

\begin{document}

\title{Coupled catastrophes: sudden shifts cascade and hop among interdependent systems}

%\author{Charles D. Brummitt, George Barnett, Raissa M. D'Souza}

\author[1,2,6]{Charles D. Brummitt\thanks{Corresponding author: \href{mailto:c.brummitt@columbia.edu}{c.brummitt@columbia.edu}}}
\author[3]{George Barnett}%\thanks{gabarnett@ucdavis.edu}}
\author[2,4,5,7]{Raissa M. D'Souza}%\thanks{raissa@cse.ucdavis.edu}}
%\affil[1]{\small Department of Mathematics, University of California, Davis, CA 95616 USA}
%\affil[2]{Complexity Sciences Center, University of California, Davis, CA 95616 USA}
%\affil[3]{Department of Communication, University of California, Davis, CA 95616 USA}
%\affil[4]{Department of Computer Science, University of California, Davis, CA 95616 USA}
%\affil[5]{Department of Mechanical Engineering, University of California, Davis, CA 95616 USA}
%\affil[6]{Santa Fe Institute, 1399 Hyde Park Road, Santa Fe, NM 87501 USA}
{\small \affil[1]{Department of Mathematics,}% University of California, Davis, CA 95616 USA}
\affil[2]{Complexity Sciences Center,}% University of California, Davis, CA 95616 USA}
\affil[3]{Department of Communication,}% University of California, Davis, CA 95616 USA}
\affil[4]{Department of Computer Science,}% University of California, Davis, CA 95616 USA}
\affil[5]{Department of Mechanical Engineering, University of California, Davis, CA 95616 USA,}
\affil[6]{Center for the Management of Systemic Risk, Columbia University, %500 W. 120th St., Mudd 510, 
New York, NY 10027 USA,}
\affil[7]{Santa Fe Institute, %1399 Hyde Park Road, 
Santa Fe, NM 87501 USA}}

\date{\today}

\begin{@twocolumnfalse}

\twocolumn[
\maketitle

\begin{onecolabstract}
An important challenge in several disciplines is to understand how sudden changes can propagate among coupled systems. Examples include the synchronization of business cycles, population collapse in patchy ecosystems, markets shifting to a new technology platform, collapses in prices and in confidence in financial markets, and protests erupting in multiple countries. A number of mathematical models of these phenomena have multiple equilibria separated by saddle-node bifurcations. We study this behavior in its normal form as fast--slow ordinary differential equations. In our model, a system consists of multiple subsystems, such as countries in the global economy or patches of an ecosystem. Each subsystem is described by a scalar quantity, such as economic output or population, that undergoes sudden changes via saddle-node bifurcations. The subsystems are coupled via their scalar quantity (e.g., trade couples economic output; diffusion couples populations); that coupling moves \new{the locations of} their bifurcations. The model demonstrates two ways in which sudden changes can propagate: they can cascade (one causing the next), or they can hop over subsystems. \newC{The latter is absent from classic models of cascades. For an application, we study the Arab Spring protests. After connecting the model to sociological theories that have bistability, we use socioeconomic data to estimate relative proximities to tipping points and Facebook data to estimate couplings among countries. We find that although protests tend to spread locally, they also seem to ``hop'' over countries, like in the stylized model; this result highlights a new class of temporal motifs in longitudinal network datasets.}

\end{onecolabstract}
]

\end{@twocolumnfalse}

\saythanks

\section{Introduction}
\newB{Sudden changes propagating among coupled systems poses a significant scientific challenge in many disciplines, yet we lack an adequate mathematical understanding of how local sudden changes spread~\cite{Barnosky:2013js}. Earth's biosphere, for example, appears to be approaching several planetary-scale sudden changes triggered by human activity, including species extinction, desertification, and lake eutrophication, that spread from one spatial patch to another~\cite{Barnosky:2013js}. That} spatial spread poses dangers but also opportunities for detecting early warning signs~\cite{vanNes:2005hl,Dai:2013ii,Dakos:2010gu}. Socioeconomic systems have examples, too: Booms and busts in business cycles in different economies appear to be synchronizing because of trade, financial, and other linkages~\cite{Bordo:2003wt,Ormerod:2002ke,Inklaar:2008hi,Calderon:2007bl}. 
\new{Poverty traps at multiple scales seem to be coupled~\cite{Barrett2006_fractal}.} Abrupt declines in an asset price can trigger sharp declines in confidence and fire sales of other assets, as occurred in the 2007--2008 global financial crisis~\cite{McCulley:2009wu}. Protests and social uprisings appear to spread contagiously among countries, with one protest \new{seeming to inspire} others via news and social media~\cite{Howard:2011vc,Hussain:2013ir}. The equilibrium supply and demand of a new technology that replaces an old one (such as compact discs replacing cassettes or electric cars replacing fuel cars) can change abruptly~\cite{Krugman1996self}, and movement of people between distinct markets can facilitate adoption of the new technology~\cite{MobileWallets}. In each of these examples, a system consists of distinct subsystems that (1) change suddenly between equilibria and  (2) are coupled. A mathematical understanding of these phenomena could pave the way to predicting and to steering these sudden changes.

\begin{table*}[htd]
\caption{Examples of coupled subsystems in which each subsystem undergoes sudden changes in the form of saddle-node bifurcations, in models cited in the column ``regime shift''. The column ``scalar quantity'' describes the state of the subsystem, and it corresponds to $x(t)$, $y(t)$, or $z(t)$ in the model in Sec.~\ref{sec:model}. Citations in the fourth column include empirical studies and mathematical models.}
\begin{center}
\begin{tabular}{p{14mm}p{35mm}p{30mm}p{67mm}}%{cccccc}
%\hline
discipline & regime shift & scalar quantity & examples of couplings among subsystems \\ \hline 
ecology & extinction due to over-harvesting~\cite{NoyMeir1975,May1977} & population & diffusion among patches of an ecosystem~\cite{vanNes:2005hl,Dakos:2010gu} \\[1mm]
economics & boom and bust in the Kaldor model of business cycles~\cite{Varian:1979fg} & output (gross domestic product) & investment between sectors~\cite{Lorenz:1987cp}, trade~\cite{Krugman1996self} and capital flows~\cite{Selover:1999bk} between countries can synchronize business cycles 
\\[1mm] 
economics & currency crisis (devaluation or, for a peg, loss of reserves)~\cite{Masson1999} & currency value & changes in macroeconomic fundamentals, sentiment, perceived riskiness, risk aversion~\cite{Masson1999}, trade~\cite{Eichengreen1996}  \\[1mm]
\new{economics} & \new{poverty trap~\cite{Azariadis2005,Matsuyama2008}} & \new{well-being (capital, capabilities)} & \new{fractal poverty traps ~\cite{Barrett2006_fractal}}   \\[1mm]
finance & asset price declines\ \cite{Gennotte1990,Brummitt2014} & asset price &  asset-to-asset contagion (a bank with a declining asset sells other assets)~\cite{Huang2013} 
\\[1mm]
finance & probability of bank failure~\cite{Ho:1980fq} & probability of bank failure & worry about institutions' creditworthiness spreads contagiously~\cite{Anand2012} \\[1mm]
technology adoption & sudden change to new platform~\cite{Krugman1996self,Herbig:1991wz} & difference between supply and demand of the new platform & movement of people among distinct markets~\cite{MobileWallets} \\[1mm]
political  & uprisings, revolts~\cite{Kuran1989,Slee2012} & number~of~protest-ors & communication spreads inspiration, successful strategies across borders~\cite{Gause2011,Howard:2011vc,Hussain:2013ir,NYT_FacebookArabSpring}; raising importance of identity~\cite{Slee2012} that span borders~\cite{Howard:2011jg}
\end{tabular}
\end{center}
\label{tab:systems}
\end{table*}%

In this paper, we take a step toward the goal of mathematically understanding how sudden changes can spread among coupled systems~\cite{Barnosky:2013js}. Our model consists of one \emph{system}, such as the global economy or a large ecosystem, that consists of multiple \emph{subsystems} coupled to one another; for example, economies of multiple countries are coupled by trade, while patches of an ecosystem are coupled by movement of organisms. To choose dynamics,  we note that many models of the aforementioned phenomena (cited in the second column of Table~\ref{tab:systems}) have one or three equilibria and an S-shaped bifurcation diagram (which is equivalent to a slice of the cusp catastrophe~\cite{Stewart:1982ws}). Thus, we let each subsystem evolve according to the normal form of this catastrophe. The state of each subsystem can change suddenly when it passes a saddle-node bifurcation, \new{one of the simplest types of ``regime shifts'' (which are sudden changes in a system's state)~\cite{Boettiger:2013br}.} Next, we introduce linear couplings between these subsystems, meaning that a change in one subsystem affects other subsystems coupled to it in proportion to that change. These couplings move the locations of the latter subsystems' bifurcations.

This model allows us to explore how regime shifts can synchronize and spread. Suppose one subsystem $\up$ ``drives'' (i.e., affects) another subsystem $\inter$, which we denote by $\up \drives \inter$. Then a regime shift in $\up$ can trigger one in $\inter$, meaning that their regime shifts synchronize. If the driven subsystem $\inter$ drives a third subsystem $\down$ (i.e., if $\up \drives \inter \drives \down$), then one possible behavior is a \emph{cascade} of regime shifts, one triggering another like falling dominoes. Another possibility is that the ``intermediate'' subsystem $\inter$ is far from its tipping point but that the others ($\up$ and $\down$) are close to their tipping points; then 
a regime shift in the driver subsystem $\up$ can nudge the intermediate subsystem $\inter$ enough to push $\down$ past its tipping point but not so much that $\inter$ passes its tipping point. That is, a sequence of regime shifts can ``hop'' over intermediate subsystems. This phenomenon is not observed in classic models of cascades (e.g., percolation, epidemic spreading, and sandpile models).

\newB{
This ``model of many models'' abstracts from many domain-specific details. It suggests what might happen in more realistic settings. To give an example, we consider protests erupting nearly simultaneously in many countries. 
We first show how two sociological theories of revolutions give rise to the same S-shaped bifurcation diagram used to model the individual subsystems of our mathematical model. We also indicate how our model can generalize these sociological theories to multiple, coupled countries in a stylized way. 
Then we consider data on the Arab Spring, the revolts and uprisings that seemingly cascaded among countries in the Middle East and Northern Africa starting in December 2010~\cite{Hussain:2013ir}. 
We explore whether protests spread locally in two networks that capture possible influence to protest, Facebook and shared borders, but we also find evidence of protests seeming to hop over countries.
}

\new{
\newA{Our approach differs} 
from the many recent studies of cascades in interdependent networks~\cite{Gao2011,Brummitt2012,Reis2014}, all of which model ``interdependence'' and ``coupling'' as occurring between pairs of nodes (individual ``agents'') belonging to different subsystems. 
Instead, we consider subsystems coupled via some aggregate quantity, such as 
investment between sectors~\cite{Lorenz:1987cp} or
the fraction of people protesting in a country~\cite{Kuran1989}. 
}

\new{Much attention is paid to regime shifts in large, central nodes, such as recessions in central economies or insolvency of large banks. Our findings suggest that small changes in these central nodes (potentially triggered by a large change in a small node adjacent to it) can suffice to trigger a regime shift in a peripheral node close to its tipping point. 
}

\section{Normal-form model of coupled subsystems with one or two stable states}
\label{sec:model}

We begin by considering two subsystems $\up$ and $\inter$, each described by a single real number, $x(t)$ and $y(t)$, that changes over time $t$. (Interpretations of $x(t),y(t)$ for various contexts are given in the third column of Table~\ref{tab:systems}.) The subsystems evolve according to the autonomous ordinary differential equations 
\begin{subequations}
\begin{align}
\frac{\diff x}{\diff t} %\equiv x' 
&= -x^3 + c x + a + \coupling{y}{x}{X} \label{eq:master_flow}   \\
\frac{\diff y}{\diff t} %\equiv y' 
&= -y^3 + d y + b + \coupling{x}{y}{Y}  \label{eq:slave_flow},
\end{align}
\label{eq:two_cubic_flows}
\end{subequations}
\!\!\!\!\!\!\!  %% I'm not sure why we need to remove horizontal space here.
where $\coupling{\cdot}{\cdot}{\cdot}$ are some coupling functions (specified later), 
and where $a, b, c, d \in \mathbb{R}$ are parameters that change slowly compared to $x(t),y(t)$, 
so System~\eqref{eq:two_cubic_flows} is a \emph{fast--slow system}~\cite{Kuehn:2011fu}. 

Variants of System~\eqref{eq:two_cubic_flows} have been studied in many contexts, including the double cusp catastrophe~\cite{Godwin:1975bh,Zeeman:1976dj,Callahan:2006vo}, cuspoidal nets~\cite{Abraham:1990wp,Abraham:1991ks}, and coupled van der Pol oscillators~\cite{Storti:1986db,Storti:1987tu,Rand:1980ub,Storti:2000uv,Pastor:1993hl,PastorDaz:1995ds,Camacho:2004ce,Low:2006jk} (for more information, see Appendix~\ref{sec:related_lit}). Coordination games and global games in economics are similar to System~\eqref{eq:two_cubic_flows} in that they also permit multiple equilibria, but they lack dynamics. Global games have been applied to currency crises~\cite{Morris1998}, debt crises~\cite{Morris2004coordination,Corsetti2006}, bank runs~\cite{Goldstein2004,Rochet2004coordination}, and riots and political change~\cite{Atkeson2000,Edmond2013}; moreover, contagion has been studied in generalizations of these models, such as currency crises triggering \newA{more} currency crises~\cite{Masson1999}, bank crises triggering \newA{more} bank crises~\cite{Goldstein2004}, and currency crises triggering bank crises~\cite{Goldstein2005}. Here we take a catastrophe-theoretic approach~\cite{Stewart:1982ws} and emphasize the role of multiple equilibria rather than eliminate multiple equilibria, as in single-period global games.

To isolate the effect of coupling, here we focus on contagion of regime shifts in a simple setting, the singular limit in which $x(t)$ and $y(t)$ change arbitrarily more quickly than the ``slow parameters'' $a,b,c,d$. Thus, we focus on the critical manifold, i.e., the solutions $(x^*,y^*)$ to System~\eqref{eq:two_cubic_flows} with $\diff x / \diff t = \diff y / \diff t = 0$.

Next we briefly review the familiar result that, in the absence of coupling, the subsystems evolving according to Eq.~\eqref{eq:master_flow}  and Eq.~\eqref{eq:slave_flow} each have two saddle-node bifurcations, and then we show how coupling functions $\coupling{\cdot}{\cdot}{\cdot}$ move those ``tipping points''.

\subsection{Uncoupled systems each undergo a cusp catastrophe}

If the coupling functions $\coupling{\cdot}{\cdot}{\cdot}$ are identically zero, then subsystems $\up$ and $\inter$ are \emph{uncoupled}, and Eqs.~\eqref{eq:master_flow} and~\eqref{eq:slave_flow} are the normal forms of the cusp catastrophe (in the special case of a minus sign on the cubic term~\cite[Theorem 8.1]{Kuznetsov2010elements}). 
We chose this form to study the general effects of couplings rather than domain-specific versions of the cusp, which are topologically equivalent to the normal form in Eq.~\eqref{eq:master_flow}. Hereafter, we take $c=d=1$ for simplicity. 

\begin{figure}[htb]
\begin{center}
\includegraphics{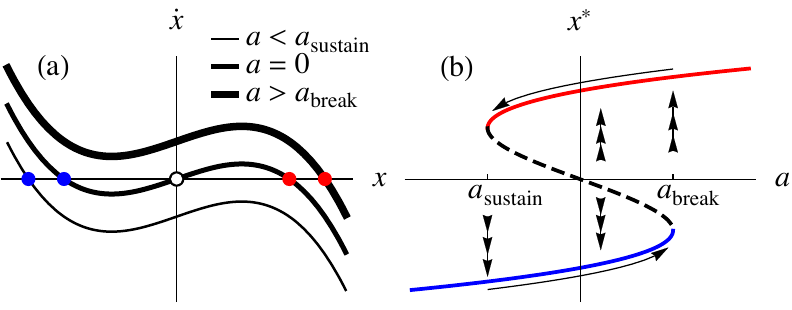}
\caption{{\bf In isolation, each system has two saddle-node bifurcations.}
(a) The flow $\dot x$ in Eq.~\eqref{eq:master_flow} has one or three equilibria, depending on the value of the parameter $a$. Filled (respectively, open) circles denote stable (unstable) equilibria. 
(b) The bifurcation diagram [the equilibria $x^*$ of~\eqref{eq:master_flow} as a function of $a$] is a slice of the cusp catastrophe, with two saddle-node bifurcations at values of $a$ denoted by $a_\text{break}$ and by $a_\text{sustain}$. The solid (respectively, dashed) curves are stable (unstable) fixed points $x^*$. 
Triple arrows denote the fast flow~\eqref{eq:master_flow}; single arrows denote a slow flow $\diff a/\diff t$ described in the text.}
\label{fig:isolated}
\end{center}
\end{figure}

If the subsystems evolving according to Eq.~\eqref{eq:master_flow} and Eq.~\eqref{eq:slave_flow} are uncoupled, then \new{both subsystems} have three equilibria for certain intervals of the slow parameters $a$ and $b$, as depicted in Fig.~\ref{fig:isolated}. In this case, the set of fixed points of Eq.~\eqref{eq:master_flow} undergoes two saddle-node bifurcations at values of $a$ that we denote by $a_\text{break}$ and by $a_\text{sustain}$ (the same terminology used in~\cite{Krugman1996self}). Each subsystem is a classic example of \emph{hysteresis}. For instance, if the equilibrium $x^*$ of Eq.~\eqref{eq:master_flow} is on the ``lower stable branch'' [the blue curve in Fig.~\ref{fig:isolated}(b)], then as $a$ increases past the ``breaking point'' $a_\text{break}$, the solution $x(t)$ jumps to the ``upper stable branch'' depicted by the red curve. [In other words, the subsystem passes a tipping point (undergoes a regime shift).] As the parameter $a$ is slowly decreased, the large equilibrium is sustained [i.e., $x(t)$ lies on the red curve] until $a$ passes $a_\text{sustain}$, at which point the subsystem $x(t)$ jumps to the lower branch.

\subsection{Master--slave with linear coupling}
\label{sec:master_slave_linear}

Next we consider the analytically-solvable case of a master--slave system with linear coupling. Specifically, subsystem $\up$ drives subsystem $\inter$ (denoted $\up \drives \inter$) according to the coupling function \new{$\coupling{x}{y}{Y} := \strength x$,} 
where the constant $\strength \in \mathbb{R}$ is the coupling strength. (For instance, consider unidirectional investment between sectors in the Kaldor business cycle model, as in~\cite{Lorenz:1987cp}, or movement of organisms from one patch of an ecosystem to another, as in~\cite{vanNes:2005hl,Dakos:2010gu}.)
Then Eq.~\eqref{eq:two_cubic_flows} becomes
\begin{subequations}
\begin{align}
\frac{\diff x}{\diff t}
&= -x^3 + x + a  \label{eq:master_flow_uncoupled}   \\
\frac{\diff y}{\diff t}
&= -y^3 +  y + b + \strength x  \label{eq:slave_flow_linear}.
\end{align}
\label{eq:master_slave_linear}
\end{subequations}
The equilibria of Eq.~\eqref{eq:master_slave_linear} can be obtained analytically by first solving for the equilibria $x^*$ of the master subsystem [Eq.~\eqref{eq:master_flow_uncoupled}], and then by using the solution(s) to calculate the equilibria $y^*$ of the slave subsystem [Eq.~\eqref{eq:slave_flow_linear}]. The saddle-node bifurcations of the slave subsystem [Eq.~\eqref{eq:slave_flow_linear}] now depend on the equilibrium value(s) $x^*$ of the master subsystem [Eq.~\eqref{eq:master_flow_uncoupled}] and on the coupling strength $\strength$; we denote \new{the slave subsystem's} bifurcations \newfinal{(with respect to $b$)} by $b_\text{break}(\strength x^*)$ and by $b_\text{sustain}(\strength x^*)$. Because $x^*$ has three possible values whenever the master parameter $a \in (a_\text{sustain},  a_\text{break})$, the slave subsystem has three possible values for each of its bifurcation points $b_\text{break}(\strength x^*)$ and $b_\text{sustain}(\strength x^*)$ whenever $a \in (a_\text{sustain},  a_\text{break})$. 

\begin{figure}[htb!]
\begin{center}
\includegraphics[width=\columnwidth]{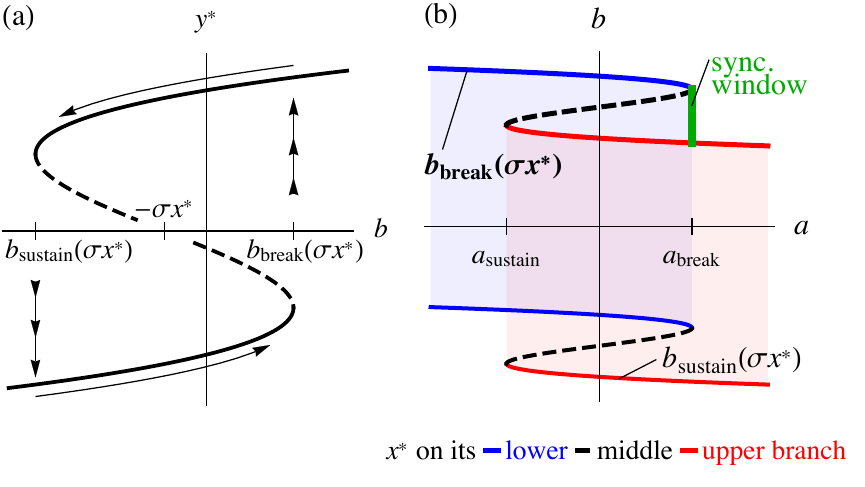}
\caption{
{\bf Coupling a slave subsystem to a master subsystem moves the slave subsystem's tipping points and can change them suddenly.} 
\new{
Panel (a): The bifurcation diagram of the slave subsystem shows the equilibrium of the slave subsystem $y^*(b; \strength x^*)$ as a function of its slow parameter $b$. (The slave subsystem's equilibrium $y^*$ also depends on the coupling term $\strength x^*$ due to the influence of the master subsystem.)  
In this example, the master subsystem has just passed its break point $a>a_\text{break}$, so the master subsystem has quickly moved to its upper stable branch of equilibria ($x^* > 0$). Because the coupling strength $\strength > 0$, the sudden shift in the master subsystem makes it easier for the slave subsystem to pass its break point [$b_\text{break}(\strength x^*) < b_\text{break}(0)]$. 
}%
Panel (b): The \new{locations of the saddle-node bifurcations} of the slave subsystem [Eq.~\eqref{eq:slave_flow_linear}], denoted by $b_\text{break}(\strength x^*)$ and by $b_\text{sustain}(\strength x^*)$, are one- or three-valued functions of $a$, the parameter of the master subsystem. The colors match those in Fig.~\ref{fig:isolated}: if the master subsystem's equilibrium $x^*$ lies on its lower (respectively, upper) stable branch depicted in Fig.~\ref{fig:isolated}(b), then the bifurcation points of the slave subsystem are the blue (respectively, red) curves in panel (b). There exist three equilibria $y^*$ in the shaded blue (respectively, red) regions. Here, $\strength > 0$, so the master subsystem acts to prevent the slave subsystem from crossing its break point $b_\text{break}(\strength x^*)$ when $x^* < 0$ and facilitates it when $x^*> 0$. If $(a,b)$ crosses the green line \new{segment} marked ``sync.\ window'', then the regime shifts \emph{synchronize}: the master subsystem [Eq.~\eqref{eq:master_flow_uncoupled}] crosses its break point $a_\text{break}$, causing $x^*$ to jump from a negative number to a positive number, which causes the slave subsystem [Eq.~\eqref{eq:slave_flow_linear}] to cross its break point $b_\text{break}(\strength x^*)$.
}
\label{fig:master_slave_linear}
\end{center}
\end{figure}

Figure~\ref{fig:master_slave_linear} shows the resulting bifurcation diagrams of the slave subsystem for $\strength = 0.1$. The saddle-node bifurcations are now \newB{functions of the coupling term:} $b_\bullet(\strength x^*) = b_\bullet(0) - \strength x^*$, where $\bullet$ is either ``\text{break}'' or ``\text{sustain}''\newB{, and $b_\bullet(0)$ is the bifurcation when the subsystems are uncoupled ($\strength = 0$)}. To understand the consequences of this displacement of the bifurcations, suppose that the coupling strength $\strength$ is positive and that the master subsystem is initially on its lower stable branch [the blue curve in Fig.~\ref{fig:isolated}(b)]. Thus,  $x(0) = x^* < 0$ and 
$\strength x^* < 0$, so the master subsystem {suppresses} a regime shift in the slave subsystem, meaning that the parameter $b$ must increase further to pass $b_\text{break}(\strength x^*)$ compared to the case of no coupling ($\strength = 0$). However, if the master subsystem passes its break point (i.e., if $a$ increases past $a_\text{break}$), then the master subsystem $x(t)$ jumps to its upper stable branch [the red curve in Fig.~\ref{fig:isolated}(b)], where $x(t) = x^* > 0$. That sudden change {facilitates} a regime shift in the slave subsystem, meaning that the parameter $b$ does not need to increase as much [in order to pass $b_\text{break}(\strength x^*) = b_\text{break}(0)-\strength x^*$] as it would if there were no coupling.

This simple system illuminates how regime shifts might {synchronize}. When the slow parameter $a$ of the master subsystem [Eq.~\eqref{eq:master_flow_uncoupled}] increases past its saddle-node bifurcation at $a_\text{break}$, the master subsystem jumps to its upper stable branch of equilibria [recall Fig.~\ref{fig:isolated}(b)], so the relevant saddle-node bifurcation for the slave subsystem [Eq.~\eqref{eq:slave_flow_linear}] suddenly changes from the blue curve to the red curve in Fig.~\ref{fig:master_slave_linear}(b). Thus, at the moment when $a$ passes $a_\text{break}$, if the value of $b$ lies above the red curve in Fig.~\ref{fig:master_slave_linear}(b) (and below the blue curve, meaning that the slave subsystem has not already jumped to its upper branch of equilibria), then the regime shift in the slave subsystem occurs \emph{simultaneously} with the regime shift in the master subsystem. The green line \new{segment} in Fig.~\ref{fig:master_slave_linear} marks the ``synchronizing window'' $S$, the interval of values of $(a,b)$ leading to synchronized regime shifts.\footnote{
\new{
Specifically, $S$ is the Cartesian product $\{a_\text{break}\} \times I$, where $I = [b_\text{break}(\strength x_\text{upper}^*),b_\text{break}(\strength x_\text{lower}^*)] \subset \mathbb{R}$ is the closed interval with the minimum (respectively, maximum) of $I$ equal to the value of $b_\text{break}(\strength x^*)$ for $x^*$ on its upper (respectively, lower) branch of equilibria at $a=a_\text{break}$. 
The ``break-type'' regime shifts synchronize if and only if the slow variables $(a,b)$ pass through $S$.
}}

For an interpretation of the synchronizing window, consider two economies $\up$ and $\inter$ that are both stuck in recession in the Kaldor business cycle model~\cite{Krugman1996self,Lorenz:1987cp}. If $\up$ undergoes a boom, does the \new{rise in the} demand of $\up$ for imports from $\inter$ push $\inter$ out of its recession? The synchronizing window $S$ specifies how close to its tipping point $\inter$ must be for the economic booms to synchronize, which provides an answer to Krugman's conjecture in~\cite{Krugman1996self}.

In summary, there are three ways in which the two subsystems in Eq.~\eqref{eq:master_slave_linear} could both pass their breaking points, $a_\text{break}$ and $b_\text{break}(\strength x^*)$. First, the slave subsystem could undergo a regime shift on its own, meaning that $b$ increases past the blue curve in Fig.~\ref{fig:master_slave_linear}(b) while $a$ remains below $a_\text{break}$, and subsequently $a$ passes $a_\text{break}$. Second, the two subsystems could simultaneously pass their breaking points, meaning that $(a,b)$ crosses the synchronizing window in Fig.~\ref{fig:master_slave_linear}(b). Third, the master subsystem could pass its breaking point $a_\text{break}$, but the slave subsystem remains too far from its tipping point (despite becoming abruptly closer), so there is a delay in time between the regime shifts.

\new{As the subsystems become more strongly coupled (larger coupling strength $\strength$),} it becomes easier for the regime shifts to synchronize: the S-shaped curves in Fig.~\ref{fig:master_slave_linear}(b) stretch vertically (but the intersections of the dashed curves and the $a=0$ axis remain fixed), so the synchronizing window $S$ enlarges with $\strength$. (For an illustration, see Fig.~\ref{fig:sync_window_vs_coupling} in the Supplement.) The results of other simple coupling functions, such as $\pm \strength |x|$, are simple transformations of Fig.~\ref{fig:master_slave_linear}(b) (see Fig.~\ref{fig:other_couplings}); we chose the coupling $\strength x$ for simplicity. The results of this subsection also apply to couplings that form a directed star graph.\footnote{A directed star coupling graph, $\{X \drives Y_i: i = 1, 2, \dots, n-1\}$, is a system with one master subsystem that evolves according to Eq.~\eqref{eq:master_flow_uncoupled} and that drives $n-1$ slave subsystems according to Eq.~\eqref{eq:slave_flow_linear}, with potentially different parameters for the various slave subsystems $\inter_i$.} \new{[A glossary of terminology for graphs and networks is provided in Appendix~\ref{sec:glossary}.]}

\newC{If the coupling were bidirectional, then the equilibria $(x^*, y^*)$ would no longer be solvable in closed form. Although synchronized regime shifts could still occur, characterizing the equilibria becomes more complicated, as illustrated by Abraham's numerical studies~\cite{Abraham:1991ks}. (For more details on related mathematical literature, see Appendix~\ref{sec:related_lit}.) Next we generalize in a way such that the equilibria remain analytically solvable.}

\subsection{Master--slave--slave system $\up \drives \inter \drives \down$}
\label{subsec:3systems}

Now we introduce a third subsystem $\down$, and we assume that $\inter$ drives $\down$ in the same way in which the master subsystem $\up$ drives $\inter$ (and with the same coupling strength $\strength$, for simplicity). Thus, we augment Eqs.~\eqref{eq:master_flow_uncoupled} and \eqref{eq:slave_flow_linear} with the equation 
\begin{align}
z' &= -z^3 + z + c + \strength y  \label{eq:slave_slave_flow_linear}
\end{align}
with a new slow parameter $c \in \mathbb{R}$ that, like $a$ and $b$, changes much more slowly than $x,y$ and $z$ do. 

Regime shifts can spread in two ways in this system $\up \drives \inter \drives \down$. 
First, if all three systems are sufficiently close to their tipping points, then a cascade of regime shifts can occur, one causing the next. The second way is more novel: if the intermediate system $\inter$ is relatively far from its tipping point whereas $\up$ and $\down$ are close to their tipping points, then the sequence of regime shifts can ``hop'' over the intermediate system $\inter$. That is, a regime shift in the master subsystem [Eq.~\eqref{eq:master_flow_uncoupled}] can nudge the intermediate system $\inter$ [Eq.~\eqref{eq:slave_flow_linear}] enough to trigger a regime shift in the third system $\down$ [Eq.~\eqref{eq:slave_slave_flow_linear}] but not so much that $\inter$ undergoes a regime shift. 

\begin{figure}[htb]
\begin{center}
\includegraphics{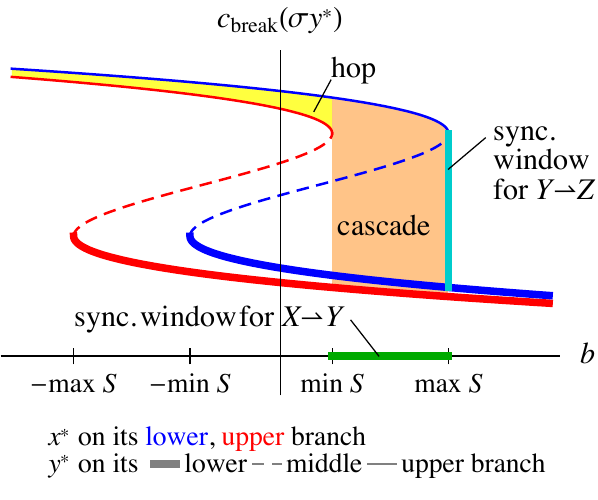}
\caption{{\bf Catastrophes can cascade, or they can hop over intermediate systems.} 
\new{The two backward-S-shaped curves are plots of the break point }
$c_\text{break}(\strength y^*)$ of the downstream system $\down$ [Eq.~\eqref{eq:slave_slave_flow_linear}] as a function of the slow parameter $b$ of the intermediate system~\eqref{eq:slave_flow_linear} for coupling strength $\strength = 0.2$. The two curves show the effect of the master parameter $a$ increasing past its break point $a_\text{break}$, at which time we change focus from the blue, right-hand curve to the red, left-hand curve. Thick curves (respectively, thin curves, dashed curves) correspond to the intermediate system $y^*$ on its upper stable branch (respectively, lower stable branch, middle unstable branch). As in Fig.~\ref{fig:master_slave_linear}(b), the green line marks the 
\new{synchronizing window $S$ for $\up$ and $\inter$ [the} values of $b$ such that, when $a$ crosses $a_\text{break}$, the intermediate system $\inter$ passes its break point $b_\text{break}(\strength x^*)$]. The cyan line marks the analogous interval for $\inter$ and $\down$ (for $a=a_\text{break}$). The orange and yellow regions are values of $(b,c)$ leading to sequences of regime shifts that cascade or that jump over $\inter$, respectively.}
\label{fig:master_slave_slave}
\end{center}
\end{figure}

We illustrate these two phenomena in Fig.~\ref{fig:master_slave_slave}, a plot of the ``downstream subsystem'' $\down$'s break point $c_\text{break}(\strength  y^*)$ at the moment when the master subsystem's parameter $a$ increases past its break point $a_\text{break}$. At this moment, the master subsystem jumps from its lower branch of equilibria to its upper branch, so we change focus from the red curve to the blue curve in Fig.~\ref{fig:master_slave_slave}. If the slow parameters $(b,c)$ lie in the orange region labeled ``cascade'' in Fig.~\ref{fig:master_slave_slave}, then a cascade of regime shifts occurs, one regime shift causing the next. To see why, note that $b$ lies in its synchronizing window $S$ [the green line in Figs.~\ref{fig:master_slave_linear}(b) and~\ref{fig:master_slave_slave}], so $\inter$ passes its break point $b_\text{break}(\strength x^*)$; and note that $c$ lies above the thick, red line $c_\text{break}(\strength y^*)$, so $\down$ passes its break point $c_\text{break}(\strength y^*)$. If, on the other hand, the parameters $(b,c)$ lie in the yellow region labeled ``hop'' in Fig.~\ref{fig:master_slave_slave}, then the sequence of regime shifts hops over the intermediate subsystem $\inter$. To see why, note that $b$ is below its synchronizing window $S$, so $b$ does not pass its break point $b_\text{break}(\strength x^*)$ when $a$ crosses $a_\text{break}$, but notice that $c$ lies above the red thin line, so $\down$ passes its break point $c_\text{break}(\strength y^*)$ despite receiving only a small nudge from $\inter$. 

Note that such ``cascade hopping'' cannot occur in many classic models of cascades, including the Ising model~\cite{Dorogovtsev2008_CriticalPhenomena}, sandpile models~\cite{Dorogovtsev2008_CriticalPhenomena}, and threshold models~\cite{Watts2002,Baxter2010}. For cascade hopping to occur, some vertices of the graph must be able to affect their neighbors in at least three ways (e.g., with a small, medium, or large amount of force). \new{A phenomenon that is qualitatively similar to cascade hopping occurs in epidemiology:} some diseases are contagious yet asymptomatic, so the sequence of contractions of the disease can appear to hop over individuals. Different people remain in the asymptomatic state for different amounts of time, which resembles coupled subsystems with different proximities to tipping points.\footnote{HIV is an example of an infection with large variability in asymptomatic periods, resulting mostly from variability among patients~\cite{HIVasymptomatic}.} Next we show circumstantial evidence that cascade hopping may occur in other kinds of contagion in human populations. 

\section{Communication-coupled outbreaks of protest}\label{sec:protest}

\new{We have presented a ``model of many models'' that captures a commonality among systems in Table~\ref{tab:systems} but that ignores many domain-specific details.  If the models in Table~\ref{tab:systems} are one step removed from reality, then the stylized model in Sec.~\ref{sec:model} is two steps removed from reality. The virtue of studying such a simple model is to elucidate what phenomena might happen in more realistic settings.} 

\newB{To give one example, \newA{in this section we consider protests and revolutions occurring in many countries.}   \newA{In Sec.~\ref{sec:sociology},} we summarize Kuran's model of protests and revolutions based on preference falsification~\cite{Kuran1989} and Slee's model of identity-driven cascades~\cite{Slee2012}. 
Our model is a stylized generalization of these models to multiple countries, with finance and cross-border identity being two possible mechanisms for coupling protests across borders. 
\newA{Next, we study data on countries involved in the Arab Spring, the uprisings in Northern Africa and in the Middle East during 2010--2011. 
Using the theoretical model of Sec.~\ref{sec:model} as a guide for asking questions, we explore the role of contagion and common cause in the Arab Spring (Sec.~\ref{sec:contagion_common_cause}), whether protests seem to spread locally (Sec.~\ref{sec:dominoes}) or in non-local jumps (Sec.~\ref{sec:hop_Arab_Spring}). }}

\subsection{\new{Models of revolutions based on preference falsification and identity}}
\label{sec:sociology}

\new{To begin, we summarize two models of protests and revolutions that emerge suddenly via saddle-node bifurcations. 
Then we explain how the conceptual framework in Sec.~\ref{sec:model} can capture a generalization of these models to multiple countries with couplings between them. (Not all models of protests have saddle-node bifurcations. For some recent examples, see~\cite{Braha2012,Lang2014,Lang2015arXiv,Berestycki2015_riotsPDE}.)

One way in which protests and revolutions can emerge suddenly is because people had been publicly declaring a preference different from their private preference~\cite{Kuran1989}. This idea, called preference falsification, has been used in several applications~\cite{KuranPreferenceFalsificationBook}. In Kuran's model of revolutions~\cite{Kuran1989}, the unit interval $[0,1]$ denotes a political spectrum, with $0$ representing the current government and $1$ representing the opposition. He assumes that people derive ``reputational utility'' from publicly declaring a certain preference in $[0,1]$, plus an ``integrity utility'' from declaring a preference close to their private preference. Slow changes in these utility functions or in the distribution of preferences can cause a large, sudden change in collective sentiment (in a saddle-node bifurcation).

Kuran's model is more rich than the model in Sec.~\ref{sec:model}, as it has utility functions, distributions of preferences, and weights of different people, but the manifold of equilibria in Kuran's model is equivalent (in a catastrophe theoretic sense~\citep{Stewart:1982ws}) to that of the isolated subsystem in Eq.~\eqref{eq:master_flow_uncoupled}. The state variable in Kuran's model is the (weighted) share of people who publicly declare that they prefer the opposition. The equilibrium~\cite[Eq.\ 8]{Kuran1989} has one or three equilibria; in the latter case, two equilibria are stable and the other unstable, as illustrated in \cite[Figures 3--7]{Kuran1989}. The difference between the thick and thin curves in Figures 3--7 of \citep{Kuran1989} is the analog of Figure~\ref{fig:isolated}(a). 

Kuran explains two ways in which a saddle-node bifurcation can occur, leaving only one equilibrium corresponding to a large public support of the opposition~\cite[Section 4.1, pages 51--53]{Kuran1989}:
\begin{enumerate}
\item a shift in the distribution of private preferences toward the opposition~\cite[Fig. 3]{Kuran1989} due to, for example, an economic downturn~\cite{Gurr_WhyMenRebel}; 
\item a change in the reputational utility terms (for example, because the opposition becomes better able to give reputational utility), causing a shift in the threshold function that marks whether someone supports the opposition or the government. 
\end{enumerate}
These two shifts correspond to changes in a ``slow variable'' [such as the variable $a$ in Equation~\eqref{eq:master_flow_uncoupled}].  

To suggest that our model might capture a generalization of Kuran's model to multiple countries, we must motivate the assumption that the state variables in different countries are coupled somehow. Kuran mentions one possible mechanism: a shift in the reputational utility (item 2 in the list above) could be ``made possible by funds provided by a foreign source''~\cite[page 53]{Kuran1989}. That is, a coupling to a foreign country (here, a financial type), could change the equations of motion such that two equilibria vanish, leaving only the equilibrium that corresponds to large support for the opposition. To continue Kuran's story, suppose that those foreign funds were sent from the opposition in a country that has just undergone massive protests, say, because that country passed a saddle-node bifurcation. This example corresponds to the master subsystem crossing its break point ($a$ passes $a_\text{break}$), and the coupling $\sigma x$ qualitatively captures the increase in the ability of the opposition in the second country to give reputational utility to supporters because of financial funds from abroad. 
 
Identity provides another possible coupling across borders. 
Gause~\cite{Gause2011} argues that pan-Arab identity is an important reason why the Arab Spring protests emerged nearly simultaneously and why it took Middle East specialists by surprise. Identity that spans borders could couple decisions to protest. For example, in Slee's model of revolutions based on rational-choice theories of identity \citep{Slee2012}, people suffer disutility due to cognitive dissonance whenever their actions differ from the norms associated with their identity. Slee considers two identities associated with the government and with the opposition. Like in Kuran's model~\cite{Kuran1989}, small changes can eliminate two equilibria, causing large protests to erupt. To continue Slee's reasoning~\cite{Slee2012}, if people protest in one country, then it becomes more important for others in a nearby country to act according to their anti-government identity. If $x(t)$ measures the share of people in one country who are protesting, then the importance of identity in the utility functions of people in a different country could vary directly with $x(t)$, such as the simple linear coupling $\coupling{y}{x}{Y} = \sigma x(t)$ studied in Sec.~\ref{sec:master_slave_linear} and Sec.~\ref{subsec:3systems}. 

These social-scientific models of revolutions based on preference falsification and identity illustrate how difficult it is to validate our coupled-threshold model with real data: these models are based on cognitive dissonance, preferences, and identity. In principle, these cognitive phenomena could be studied with surveys, ethnographies, and other labor-intensive methods. These theories~\citep{Kuran1989,Slee2012}, which are grounded in social scientific understand of human behavior, can be seen as the connection between our conceptual model and real systems. When we describe our model as a ``model of models'' and hence two steps removed from reality, we have in mind models like~\cite{Kuran1989,Slee2012} that have bistability.
}

\new{Multiple equilibria can also arise if people have greater incentives to protest as more people decide to protest (i.e., strategic complementarities)~\cite{Granovetter1978}, for example because of safety in numbers~\cite[page 18]{Kuran1991}. Multiple equilibria also occur in repeated coordination games in which people learn about the number of protestors needed to overthrow the regime (so-called dynamic global games)~\cite{Angeletos2007}.}

\new{Now that we have connected the stylized model in Sec.~\ref{sec:model} and sociological literature such as~\cite{Kuran1989,Slee2012}, we next investigate data from the Arab Spring with questions generated from the conceptual framework of Sec.~\ref{sec:model}.}

\subsection{Contagion versus common cause in the Arab Spring}
\label{sec:contagion_common_cause}

Why did many protests begin nearly simultaneously in the Arab Spring? One explanation is \emph{common cause} (called the \emph{monsoonal effect} in the context of contagious currency crises~\cite{Masson1999}): an external driver, such as rising global food prices, pushes all countries past their tipping points (as suggested in~\cite{Lagi2011}). Another explanation is \emph{contagion}: couplings among countries (such as communication) helped to synchronize their protests. 
The analog of common cause in System~\eqref{eq:two_cubic_flows} in Sec.~\ref{sec:model} is that the slow parameters $a$ and $b$ both increase and pass their tipping points simultaneously (or at nearly the same time), with or even without coupling. The analog of contagion is that the slow parameter $a$ increases past its tipping point, which (via the coupling) pushes $b$ past its tipping point. 

To begin to explore the possible roles of common cause and contagion, we study data on attributes of countries~\cite{WorldBankData,FoodPriceData,Hussain:2013ir} and data on communication between countries via Facebook and telephone. \new{Communication across borders spread inspiration to protest, freedom memes, and strategies for success~\cite{Howard:2011vc,Hussain:2013ir,NYT_FacebookArabSpring}. Therefore, cross-border communication via Facebook and telephone may have spurred people  to publicly declare their private preferences~\cite{Kuran1989} or to act according to norms of their government-opposing identities~\cite{Slee2012}.
}
The Facebook data available~\cite{FacebookData} is coarse-grained: for each country, we have the ranked list of the top five other countries to which members of the focal country have the most friends (in 2012, the only year available \newB{to us}).

\begin{figure*}[htb]
\begin{center}
\includegraphics[scale= 1.1]{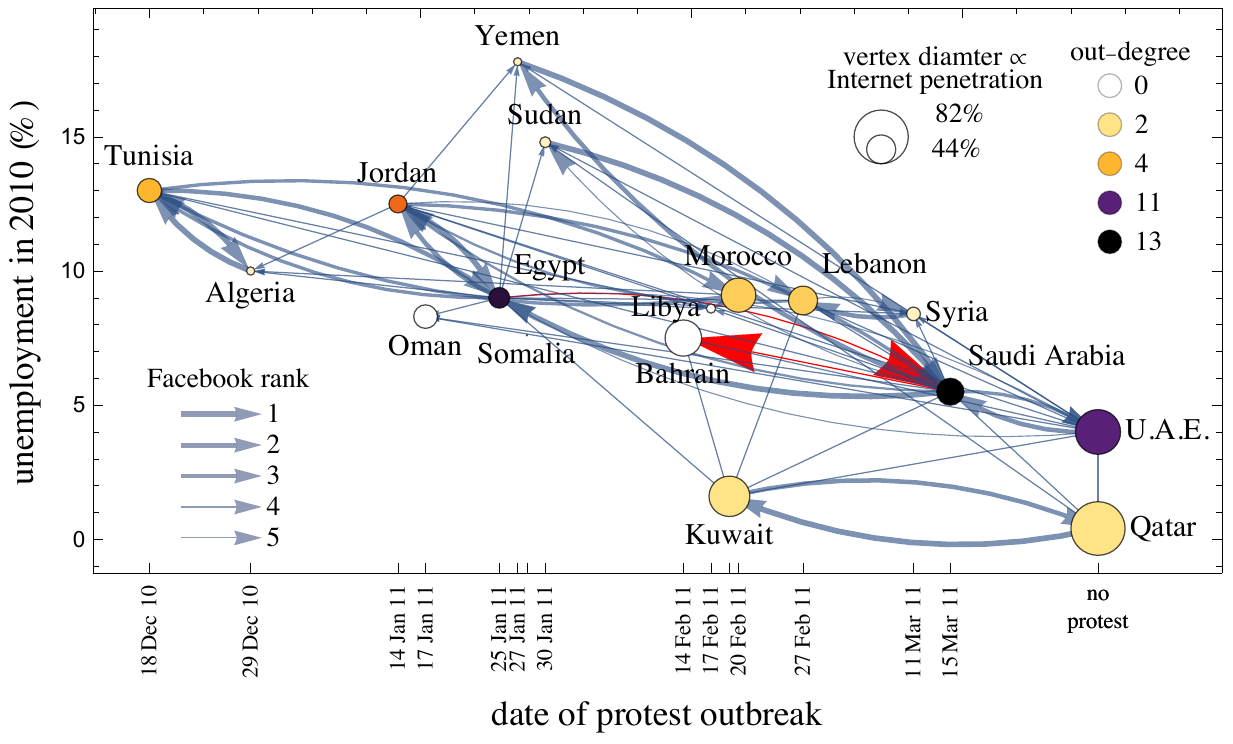}
\caption{
{\bf Exploration of the roles of proximities to tipping points and coupling (cross-border communication) for countries involved in the Arab Spring.} 
Shown is the unemployment in 2010~\cite{WorldBankData} versus the date at which protests began~\cite{WikipediaArabSpring} (see Table~\ref{tab:dates} in the Supplement) for $16$ countries %with 
\new{that had} protests, plus Qatar and \new{the} U.A.E. The best-fit line has slope $-0.5\%$ per week ($p$-value 0.06, $R^2 = 0.25$; U.A.E.\ and Qatar ignored). The direction of the Facebook edges~\cite{FacebookData} captures the spread of influence to protest:  
an edge from $i$ to $j$ means that $i$ is among $j$'s five countries with which $j$ has the most \new{Facebook} friends (in the year 2012); edge thickness decreases linearly with rank, so the thickest edges correspond to rank one (the strongest coupling). Vertex color denotes out-degree; high out-degree nodes (e.g., Egypt and Saudi Arabia) may be particularly influential in spreading influence to protest via Facebook. Highlighted in red edges with large arrowheads is one cascade hop motif, Egypt $\rightarrow$ Saudi Arabia $\rightarrow$ Bahrain (see Definition~\ref{def:hop} for the definition and Table~\ref{tab:hopmotifs} for the other nine hop motifs). A version of this plot with a few more countries that protested much later than the dates shown here is in Fig.~\ref{fig:Facebook_all} in the Supplement.
}
\label{fig:Facebook}
\end{center}
\end{figure*}

Figure~\ref{fig:Facebook} shows the subgraph of this Facebook graph induced by countries that protested in the Arab Spring, together with two countries that did not protest but that may have communicated influence to protest \new{and that shared Arab identity}, Qatar and the United Arab Emirates.\footnote{\new{Expatriates in other countries may be important, too, but we add only Qatar and the United Arab Emirates to our analysis given their geographic and cultural proximity to the countries with protests.}} 
Next, to explore the possible roles of common cause and contagion, we study what attributes of countries correlate with when their protests began.
We found that unemployment most significantly correlates with protest start date (see the downward trend in Fig.~\ref{fig:Facebook}). That suggests that high-unemployment countries were closer to their tipping points.\footnote{Linear regression indicates that each additional $1\%$ in unemployment in 2010 is associated with protests starting $3.2 \pm 1.6$ days earlier (mean $\pm$ 1 standard error); the $p$-value is 0.06 and $R^2$ is $0.25$, suggesting statistical significance.}

Internet penetration, the fraction of the population that uses the Internet~\cite{WorldBankData}, which is plotted as vertex diameter in Fig.~\ref{fig:Facebook}, may indicate the strength of coupling to other countries via social media such as Facebook, which was thought to be an important channel for inspiring protests~\cite{Howard:2011vc, Howard:2011jg}. However, Internet penetration is a weak and statistically insignificant predictor of when protests started in various countries.\footnote{Each percentage of Internet users delays the protest start date by $0.44 \pm 0.37$ days; $p$-value $0.25$, $R^2=0.099$}

Spikes in commodity food prices have been proposed as a significant cause of the Arab Spring~\cite{Lagi2011}. Here we consider consumer food prices~\cite{FoodPriceData}, which did not noticeably spike in 2010~\cite{NewSecurityBeat2014}; we found that these indices in 2010 were not predictive of when protests began in different countries.\footnote{The index~\cite{FoodPriceData} is expressed on a scale such that it equals $100$ in the year $2000$. Each unit increase in the consumer food price index is associated with protests occurring $0.09 \pm 0.26$ days earlier, but with $p$-value $0.74$ and $R^2 = 0.011$.}

Many other covariates, from economic indicators to political freedoms, were similarly weak and statistically insignificant predictors of when protests began (for the list of covariates, see the first column of Figs.~\ref{fig:motif:economic} and~\ref{fig:motif:social}). Furthermore, using the criterion for forward selection, we could not reject the null hypothesis that any of these covariates could be considered together with unemployment. 

Because this network data is longitudinal, hazards models~\cite{NetCox} or generalized estimating equations~\cite{Christakis2013} could be useful. A challenge, however, is the small sample size (about a dozen countries).

\subsection{\new{Did Arab Spring protests spread locally?}} 
\label{sec:dominoes}

\new{
The ``domino hypothesis''---that the Arab Spring protests spread locally like falling dominoes---has been the subject of speculation~\cite{Bertrand2011,Hussain:2013ir,Ellenberg2015} but\newB{, to our knowledge,} little analysis. An alternative hypothesis, motivated by the ``hopping'' phenomenon in Sec.~\ref{subsec:3systems}, is that Arab Spring protests spread non-locally in some way.

We find circumstantial evidence in support of both hypotheses. In support of the domino hypothesis, 
we found that, among countries that had protests, a majority of those countries share a border\footnote{\new{We assume that Bahrain borders Saudi Arabia, Qatar, and the U.A.E. given the proximity and the fact that Saudi Arabia and the U.A.E. sent troops to Bahrain to quell protests there~\cite{SAUAEsendTroopsToBahrain}, suggesting that protests Bahrain would likely spread to Saudi Arabia and the U.A.E.}} with at least one country whose protests began earlier, and a majority have a Facebook link from at least one country whose protest began earlier. These results are statistically significant compared to a null model of randomized protest dates.\footnote{
\new{Specifically, $7$ of the $16$ countries with protests in Fig.~\ref{fig:Facebook} and $11$ of all $20$ countries with protests (listed in Table~\ref{tab:dates}) shared a border with at least one country with protests; comparing this outcome to randomized protest start times gives a $p$-value of $0.02$ in both cases. For the Facebook graph, $10$ out of the $16$ countries with protests in Fig.~\ref{fig:Facebook} and $14$ out of all $20$ countries with protests (listed in Table~\ref{tab:dates}) 
had at least one incoming Facebook link to a country with protests that started earlier ($p$-values $0.06, 0.16$, respectively). 
}} 
In addition to this evidence of local spread, we also find evidence of protests spreading non-locally, discussed next.
}

\subsection{Cascade hopping in the Arab Spring}
\label{sec:hop_Arab_Spring}

\new{
Here we show circumstantial evidence that protests  may have spread in a non-local way consistent with the ``cascade hopping'' phenomenon in Sec.~\ref{subsec:3systems}. An empirical signature of the ``hopping'' phenomenon---though not conclusive evidence of it---is a small subgraph in which protests appear to hop over a country. If this small subgraph appears more often compared to a reasonable null model, then this subgraph is called a ``motif''. We call this particular motif a ``hop motif'' and define it as follows. (For definitions of network terminology, see the Glossary in Appendix~\ref{sec:glossary}.) 
}

\begin{definition}\label{def:hop} A \emph{hop motif} in a directed coupling graph is a triple of countries $(\up, \inter, \down)$ such that 
\begin{enumerate} 
\item the subgraph induced by $\{\up, \inter, \down\}$ is the directed path $\up \drives \inter \drives \down$;\footnote{For the Facebook graph~\cite{FacebookData}, this path subgraph means that $\up$ is on $\inter$'s top-$5$ list (of countries to which people in $\inter$ have the most friends) and that $\inter$ is on $\down$'s top-$5$ list.} \item there is no coupling edge \new{pointing} to $\down$ from any country that began to protest before $\down$ did; \item protests began first in $\up$, then in $\down$, and then in $\inter$ (or $\inter$ did not have any protests). \end{enumerate} 
\end{definition}

For the Facebook network shown in Fig.~\ref{fig:Facebook}, ten \new{triples of countries}, 
listed in Table~\ref{tab:hopmotifs}, satisfy these criteria in Definition~\ref{def:hop}.  One of them, Egypt $\rightarrow$ Saudi Arabia $\rightarrow$ Bahrain, is highlighted with red edges and large arrowheads in Fig.~\ref{fig:Facebook}. \new{Compared to a null model with random protest start times, the network in Fig.~\ref{fig:Facebook} has more hop motifs than $93.3\%$ of randomized versions.}

\new{These hop motifs suggest (but do not conclusively show) that Saudi Arabia and Egypt played the role of an intermediate subsystem $\inter$ in Sec.~\ref{subsec:3systems}. Specifically, the motifs suggest that  Saudi Arabia and Egypt may have received influence from protesting countries that played the role of the upstream subsystem $\up$ (e.g., Tunisia, Jordan) and propagated influence to other countries that played the role of the downstream subsystem $\down$ (e.g., Bahrain, Oman), which may have helped to trigger protests in $\down$ before protests began in $\inter$. Consistent with relative deprivation theory (which argues that economic stress puts countries close to a tipping point)~\cite{Gurr_WhyMenRebel}, we find that the upstream and downstream countries in Table~\ref{tab:hopmotifs} were relatively closer to tipping points than intermediate countries (see Sec.~\ref{sec:hop_motif_properties} of the SM). }

Unlike work on temporal motifs in telephone call data~\cite{Kovanen2011,Kovanen2013}, here events occur on the nodes rather than on the edges (i.e., protests occur in countries, whereas phone calls occur between individuals).  Thus, hop motifs were not studied in work on temporal motifs~\cite{Kovanen2011,Kovanen2013}. 

Note that a hop motif $(\up, \inter, \down)$ is delicate: 
\new{a communication link from $\up$ to $\down$ could explain why protests began in $\down$ before they began in $\inter$.} \new{None of the upstream and downstream countries $\up$ and $\down$ share a border, and only Jordan and Oman had a significant amount of cross-border telephone calls in 2010 ($8.3 \times 10^6$ minutes), which eliminates two of the ten hop motifs in the Facebook network (Table~\ref{tab:hopmotifs}).} Data on other communication between countries, such as cross-border mentions of hashtags on Twitter~\cite{Howard:2011vc} and consumption of news media, could reveal communication from $\up$ to $\down$, but obtaining such data is difficult and beyond the scope of this paper. \new{A limitation of the Facebook dataset is that we only know the top-$5$ countries to which each country has the most Facebook friends; considering only the top-$R$ lists with $R \in \{1,2,3,4\}$ did not result in any new hop motifs.}

\begin{table}[htdp]
\caption{Ten ``hop motifs'' in the Facebook data (see Definition~\ref{def:hop}). \new{The notation ``$\up \xrightarrow{r} Y$'' means that country $\up$ is located at position $r$ on the list of countries ranked in descending order by the number of Facebook friends with people in country $\inter$. For example, ``Egypt $\xrightarrow{1}$ Saudi Arabia'' means that Saudis have more Facebook friends in Egypt than in any other country.}}
\begin{center}
\begin{tabular}{c}
 \text{upstream $\up$} $\xrightarrow{r}$ \text{intermed.\ $\inter$} $\xrightarrow{r}$ \text{downstream $\down$} \\
 \hline
 \text{Egypt} $\xrightarrow{1}$ \text{Saudi Arabia} $\xrightarrow{1}$ \text{Bahrain} \\
 \text{Yemen} $\xrightarrow{2}$ \text{Saudi Arabia} $\xrightarrow{1}$ \text{Bahrain} \\
 \text{Tunisia} $\xrightarrow{3}$ \text{Egypt} $\xrightarrow{1}$ \text{Jordan} \\
 \text{Jordan}  $\xrightarrow{2}$ \text{Egypt} $\xrightarrow{4}$ \text{Oman} \\
 \text{Tunisia} $\xrightarrow{3}$ \text{Egypt} $\xrightarrow{4}$ \text{Oman} \\
 \text{Egypt} $\xrightarrow{3}$ \text{Kuwait} $\xrightarrow{4}$ \text{Bahrain} \\
 \text{Jordan} $\xrightarrow{4}$ \text{Saudi Arabia} $\xrightarrow{1}$ \text{Bahrain} \\
 \text{Jordan} $\xrightarrow{4}$ \text{Saudi Arabia} $\xrightarrow{2}$ \text{Oman} \\
 \text{Sudan} $\xrightarrow{5}$ \text{Saudi Arabia} $\xrightarrow{1}$ \text{Bahrain} \\
 \text{Egypt} $\xrightarrow{5}$ \text{U.A.E.} $\xrightarrow{2}$ \text{Bahrain} \\
\end{tabular}
\end{center}
\label{tab:hopmotifs}
\end{table}%

\section{Discussion}\label{sec:discussion}

Some of the most pressing global challenges involve the prediction and control of sudden changes propagating among coupled subsystems, such as avoiding disastrous shifts in the biosphere~\cite{Barnosky:2013js} and preventing crises in the financial system~\cite{McCulley:2009wu}. Livelihoods could also improve if sudden adoption of technologies in coupled markets were facilitated~\cite{Krugman1996self, MobileWallets}, or if coupled recessions and booms in economies were better managed~\cite{Krugman1996self,Bordo:2003wt,Ormerod:2002ke,Inklaar:2008hi,Calderon:2007bl}, or if social uprisings spreading among countries were better forecast~\cite{Gause2011,Howard:2011vc,Hussain:2013ir}. Mathematically understanding tipping points in coupled subsystems is a step toward meeting these challenges.

In this paper, we have shown \new{in a conceptual model} how regime shifts can propagate among coupled subsystems by cascading or even by jumping over subsystems. Here, we model a regime shift as a parameter passing a saddle-node bifurcation, which causes a sudden change to a different equilibrium. Such behavior appears in many systems~\cite{Scheffer2009,Scheffer2012} but is not the only kind of regime shift~\cite{Boettiger:2013br,Hu2014}. 
\new{This model combines continuous and discrete, threshold-like changes. The study of models with these features is a challenge in several disciplines, such as in failures spreading in economic input--output models~\cite{Baqaee2015} or in electric power grids (though in a more non-local way)~\cite{Hines2010,Eppstein2012}. We also find non-local spread: the next subsystem to pass a tipping point may lie two or more ``hops'' away from those that have passed their tipping points. 
}

\new{This model captures just one aspect of many models (couplings and saddle-node bifurcations), but it ignores many domain-specific details that could also be quite important. At best, this  ``model of many models'' can suggest what phenomena might occur in more complicated, domain specific models or in real data. As an example, we} find ten ``hop motifs'' (i.e., sequences of sudden changes that appear to hop over intermediate subsystems) in data on communication among countries involved in the Arab Spring protests. 

Much 
attention is devoted to regime shifts in large, central nodes, such as the effect of recessions in large economies or the question of whether to bail out large banks. Our findings suggest that small, seemingly innocuous changes in these central nodes (perhaps triggered by a large change in a small node adjacent to it) can suffice to trigger a regime shift in a peripheral node close to its tipping point. Such dynamics may have occurred in the aftermath of the 2008 financial crisis given that in the United States hundreds of small banks failed but few large banks failed~\cite{SmallBankFailures}. Peripheral players in networks may be vulnerable to sequences of regime shifts that hop over the core, an issue that seems to merit further attention. 

An open challenge is to estimate tipping points (if they exist at all) in various complex systems, using data from historical examples. Considering data not only from the Arab Spring but also from other episodes of nearly synchronous uprisings (e.g., in Soviet countries in 1989~\cite{Kuran1991,Lohmann1994,Kern2011} and others~\cite{Muller1986}) could elucidate how couplings among countries affect their proximities to tipping points. This understanding could enable better prediction of the next protest or revolt, complementing new techniques for mining news for sentiment and tone~\cite{Leetaru2011,Ball2011} and early warning signals applied to social network activity~\cite{Kuehn2014}. Similar advances have been made in understanding contagion of currency and debt crises among countries in the 1990s~\cite{Masson1999}. 

Another challenge is to extend work on temporal motifs in telephone call data~\cite{Kovanen2011,Kovanen2013} to settings like the one considered here. In the systems summarized in Table~\ref{tab:systems} and in the model in Sec.~\ref{sec:model}, events occur on the nodes (rather than on the edges~\cite{Kovanen2011,Kovanen2013}), and nodes can be influenced by multiple ongoing events (rather than participating in just one event at a time~\cite{Kovanen2011,Kovanen2013}). Hop motifs are just one example in this new class of temporal motifs.

\section*{Author Contributions}
\newfinal{C.D.B. conceived and carried out the study, performed empirical analysis, and wrote the paper; G.B. contributed data and helped interpret its analysis; R.M.D. helped design the study and write the paper. All authors gave final approval for publication.}

\section*{Acknowledgments}
The authors thank Kartik Anand, Prasanna Gai, and Mason Porter for suggesting references. 

\section*{Funding}
C.D.B. was supported by the Department of Defense (DoD) through the National Defense Science \& Engineering Graduate Fellowship (NDSEG) Program and by the James S. McDonnell Foundation through the Postdoctoral Fellowship Awards in Studying Complex Systems. This work was supported in part by NSF Grant No. ICES-1216048; the U.S. Army Research Office MURI Award No.\ W911NF-13-1-0340 and Cooperative Agreement No.\ W911NF-09-2-0053; the Defense Threat Reduction Agency Basic Research Award HDTRA1-10-1-0088; and The Minerva Initiative, Grant No.\ W911NF-15-1-0502.

\section*{Data accessibility}
The following data used in this article are freely available online: 
\begin{itemize}
\item Unemployment in 2010, unemployment of young men (age 15--24) in 2010, GDP per capita in 2010 (based on purchasing power parity in constant 2011 US dollars), and Internet penetration in 2010 are from The World Bank~\cite{WorldBankData}.
\item GDP per capita for Djbouti, Libya, Syria, and Somalia were missing in the World Bank data~\cite{WorldBankData}. To fill these gaps, we used the GDP per capita (for the year 2011) in the Wolfram|Alpha knowledgebase (\url{http://www.wolframalpha.com/}).
\item Missing data on Internet penetration in 2010 for Somalia in the World Bank data~\cite{WorldBankData} was filled using data from~\cite{InternetUseData} by linearly interpolating between the Internet use in 2009 ($1.16\%$) and 2011 ($1.25\%$), arriving at the estimate of $1.2\%$.
\item The Gini coefficient of the wealth distribution, level of oil production 
%(labeled ``fuel'' in Fig.~\ref{fig:motifproperties}), 
(Fig.~\ref{fig:motif:economic}), 
fraction of the population living in urban areas (Fig.~\ref{fig:motif:social}),  %(labeled ``urban'' in Fig.~\ref{fig:motifproperties}), 
fraction of the population under age $25$ (Fig.~\ref{fig:motif:social}),  % (labeled ``youth'' in Fig.~\ref{fig:motifproperties}), 
and success of the protest in achieving its goals with minimal violence (Fig.~\ref{fig:motif:social}),  %(labeled ``success'' in Fig.~\ref{fig:motifproperties}) 
are from Ref.~\cite[Table 1]{Hussain:2013ir}.
\item Political freedom data (the fourth and fifth rows of Fig.~\ref{fig:motif:social}) are from Freedom House~\cite{PoliticalFreedomData}.
\item Freedom of the press data (the last row of Fig.~\ref{fig:motif:social}) is from~\cite{PressFreedomData}.
\item The Facebook data~\cite{FacebookData} was scraped from a Facebook blog post. 
\item The cross-border telephone data~\cite{TelephoneData} is from 2010. For each country, the data has the top $10$ to $20$ countries with the most outgoing telephone calls from the focal country, measured in millions of minutes. We only have data for pairs with at least a million minutes. This data was purchased from TeleGeography, and it is available at \url{http://spins.ucdavis.edu/}. 
\item Table~\ref{tab:dates} gives the dates at which protests, demonstrations or conflicts began in the twenty countries that had some form of demonstration or conflict during the Arab Spring (from~\cite{WikipediaArabSpring}). These dates are the horizontal positions of countries in Fig.~\ref{fig:Facebook} in the main text and in Fig.~\ref{fig:Facebook_all}. 

\begin{table}[htb]
\caption{Dates when protests, demonstrations, or conflicts began in the Arab Spring.}
\begin{center}
\begin{tabular}{ll}
country & date protests began \\
\hline
\text{Tunisia} & \text{December 18, 2010} \\
 \text{Algeria} & \text{December 29, 2010} \\
 \text{Jordan} & \text{January 14, 2011} \\
 \text{Oman} & \text{January 17, 2011} \\
 \text{Egypt} & \text{January 25, 2011} \\
 \text{Yemen} & \text{January 27, 2011} \\
 \text{Djibouti} & \text{January 28, 2011} \\
 \text{Somalia} & \text{January 28, 2011} \\
 \text{Sudan} & \text{January 30, 2011} \\
 \text{Bahrain} & \text{February 14, 2011} \\
 \text{Libya} & \text{February 17, 2011} \\
 \text{Kuwait} & \text{February 19, 2011} \\
 \text{Morocco} & \text{February 20, 2011} \\
 \text{Mauritania} & \text{February 25, 2011} \\
 \text{Lebanon} & \text{February 27, 2011} \\
 \text{Syria} & \text{March 11, 2011} \\
 \text{Saudi Arabia} & \text{March 15, 2011} \\
 \text{Israel} & \text{May 15, 2011} \\
 \text{Palestinian Territory} & \text{September 04, 2012} \\
 \text{Iraq} & \text{December 23, 2012} \\
\end{tabular}
\end{center}
\label{tab:dates}
\end{table}%

\item Shared borders were computed using the function \textbf{CountryData} in the Wolfram Language~\cite{Mathematica10p1}. 
\end{itemize}

\appendix

\section{Glossary of network terms}
\label{sec:glossary}

\newfinal{
\begin{itemize}
\item A \emph{network} (or \emph{graph}) is a collection of \emph{nodes} (or \emph{vertices}) and a list of connections (or \emph{edges}) among them. 
For example, in a social network, the nodes are people and the connections could be friendships. 
%For example, the nodes could be countries, and the edges could be the list of who 
 If those connections have a direction, then the graph is called \emph{directed}; otherwise the graph is called \emph{undirected}. Graphs are typically visualized by drawing the nodes as circles and the edges among them as lines; if the edges are directed, then the lines have arrowheads to indicate their direction. %(with arrowheads on the lines if they are directed). %as circles (nodes) with lines or arrows drawn between them (for undirected and directed edges, respectively). % edges are typically drawn with arrows.
 \item A node's \emph{degree} is the number of connections it has. 
 A node in a directed graph has
an \emph{in-degree} and an \emph{out-degree}, which are the numbers of incoming and outgoing connections, respectively. 
\item A \emph{subgraph} of a graph is a graph that is entirely contained in the original graph. A \emph{subgraph induced by a certain subset of nodes} is the subgraph consisting of all the edges among those nodes. 
\item A \emph{motif} of a graph is a small subgraph that appears rather frequently compared to some randomized version of the graph.
\end{itemize}
}

%\new{
%
%\begin{itemize}
%\item A \emph{network} (or \emph{graph}) $G = (V,E)$ is a set of \emph{nodes} (or \emph{vertices}) $V$ and a set of \emph{edges} (or \emph{links}) $E \subseteq V \times V$, which is a list of pairs of nodes. 
%\item A \emph{subgraph} of a graph $G = (V,E)$ is a graph $G_0=(V_0,E_0)$ that is contained in $G$, meaning that $V_0$ is contained in $V$ and $E_0$ is contained in $E$. 
%\item A \emph{subgraph of $G = (V,E)$ induced by a set of nodes} $V_0 \subseteq V$ is the graph with node set $V_0$ and edge set $E_0 \subseteq E$ containing all edges $e$ in $E$ such that both vertices in $e$ belong to $V_0$. 
%\item A \emph{motif} of a graph is a small subgraph. Usually, a motif is a subgraph that appears rather frequently compared to some randomized version of the graph.
%\end{itemize}
%}

\section{Literature related to System~\eqref{eq:two_cubic_flows}}
\label{sec:related_lit}
%
%%%%%%%%%%%%
%
%
% double cusp catastrophe
%
%
%%%%%%%%%%%%
%
The case of System~\eqref{eq:two_cubic_flows} 
with bidirectional, symmetric coupling 
\new{ $\coupling{y}{x}{X} = \strength y, \coupling{x}{y}{Y} = \strength x$ }
is a special case of the double cusp catastrophe, which is given by the potential \begin{align*}
F &= -x^4 - y^4 + a_{22} x^2 y^2 + a_{12}x y^2 + a_{21} x^2 y \\
&\quad + c x^2 + d y^2 + \strength xy + a x + by.
\end{align*}
This singularity has very rich structure~\cite{Godwin:1975bh,Zeeman:1976dj,Callahan:2006vo}.

%
%%%%%%%%%%%%
%
%
% cuspoidal nets
%
%
%%%%%%%%%%%%
%
R.\ Abraham et al.~\cite{Abraham:1991ks} numerically studied a system similar to the double catastrophe, namely $x' = -x^3 + \bar b x + \bar a y, y' = -y^3 + \bar d y + \bar c x$, which is System~\eqref{eq:two_cubic_flows} 
but with no constant terms (i.e., $a=b=0$) and with coupling functions 
%$\cpl{x}{y} = ay, \cpl{y}{x} = cx$ 
\new{$\coupling{y}{x}{X} = \bar a y, \coupling{x}{y}{Y} = \bar c x$ }
providing the only terms independent of $x$ and independent of $y$, respectively. They numerically study the bifurcation sets by plotting the number of equilibria as a function of the four parameters \new{$(\bar a, \bar b, \bar c, \bar d)$}. 
R.\ Abraham~\cite{Abraham:1990wp} also outlined how one might study this system with $n$ equations coupled via some graph; our paper can be seen as an implementation of this idea. 

%
%%%%%%%%%%%%
%
%
% van der Pol oscillators
%
%
%%%%%%%%%%%%
%
The widely studied van der Pol oscillator $\ddot u - \mu (1-u^2) \dot u + u = 0$, upon a Li\'enard transformation 
$v = u - u^3 / 3 - \dot u / \mu$, becomes 
\begin{subequations}
\begin{align}
\dot u &= \mu (-x^3 / 3 + x - v) \label{eq:VdP1}\\
\dot v &=  u /\mu \label{eq:VdP2}.
\end{align}
\end{subequations}
Note that Eq.~\eqref{eq:VdP1} has the same form as Eq.~\eqref{eq:master_flow} in the uncoupled case 
\new{[$\coupling{\cdot}{\cdot}{\cdot} \equiv 0$].} 
Equation~\eqref{eq:VdP2} is a differential equation for the parameter that plays the role of $a$ in Eq.~\eqref{eq:master_flow}. 
The van der Pol oscillator has a unique, stable limit cycle around the origin~\cite{Storti:1986db}. Many papers have studied coupled van der Pol oscillators, with a focus on stability of oscillations~\cite{Storti:1986db,Storti:1987tu,Rand:1980ub,Storti:2000uv} and on chaos~\cite{Pastor:1993hl,PastorDaz:1995ds}, many inspired by biological applications~\cite{Rand:1980ub,Storti:1986db,Camacho:2004ce,Low:2006jk}. A related limit-cycle oscillator is the Fitzhugh--Nagumo model, a two-dimensional ODE that, when coupled to another such system, can produce chaos~\cite[Sec.\ 6.3.3]{Sprott2010elegant}.
Here, we focus on the contagion of regime shifts in the singular limit, which corresponds to the limit $\mu \to \infty$ in Eq.~\eqref{eq:VdP2}. 
%
%%%%%%%%%%%%
%
%
% end of footnote
%
%
%%%%%%%%%%%%
%

%\bibliographystyle{abbrv}
%\bibliographystyle{plainurl}
%\bibliographystyle{plain}
%\bibliographystyle{plainnat}
\bibliographystyle{abbrvnat}

%{%\small 
%\bibliography{Bibliography_CoupledCatastrophes}
%
%
%}

\clearpage

\twocolumn[\centering \LARGE  {\bf Supplementary Material\\\vspace{1cm}}]

% For section headers starting with S
\setcounter{figure}{0}
\setcounter{section}{0}
\makeatletter 
\renewcommand{\thefigure}{SM-\@arabic\c@figure}
\renewcommand{\theequation}{SM-\@arabic\c@equation}
\renewcommand{\thetable}{SM-\@arabic\c@table}
\renewcommand{\thesection}{SM-\@arabic\c@section}
\renewcommand{\thesubsection}{SM-\@arabic\c@subsection}
\makeatother

%\appendix
%\begin{appendices}

\section{Stronger coupling makes it easier to see synchronized regime shifts}

\begin{figure*}[htb]
\begin{center}
\includegraphics{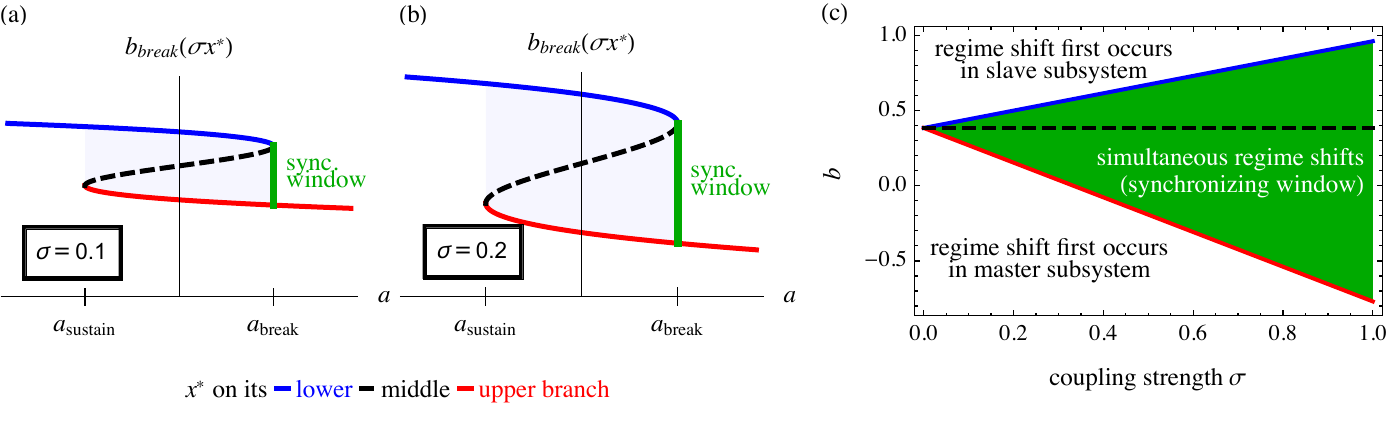}
\caption{{\bf Stronger coupling facilitates synchronization of regime shifts.} 
When $a$ increases past $a_\text{break}$, a regime shift occurs simultaneously in the master--slave system~\eqref{eq:master_slave_linear} if and only if $b$ [which is the slow parameter of the slave subsystem, Eq.~\eqref{eq:slave_flow_linear}] lies in the synchronizing window $S = \left [\frac{2}{3 \sqrt 3} \left (1 - 3 \strength \right ), \frac{2}{3 \sqrt 3} \left (1 + 3 \strength \right ) \right ]$, depicted as a green line in panels (a) and (b) and as a green region in panel (c). Panels (a) and (b) show how the S-shaped curves $b_\text{break}(\strength x^*)$ [i.e., the ``break point'' saddle-node bifurcation of the slave subsystem, Eq.~\eqref{eq:slave_flow_linear}] stretch vertically as the coupling strength $\strength$ is increased [$\strength = 0.1$ in panel (a) and $\strength = 0.2$ in panel (b)], thereby enlarging the synchronizing window. Panel (c) shows the synchronizing window $S$ as a function of the coupling strength $\strength$. If $b$ is below (respectively, above) $S$ when $a=a_\text{break}$, then the regime shift first occurs in the master (respectively, slave) subsystem, and there is a delay between regime shifts. The dashed line marks the value of $b_\text{break}(\strength x^*)$ for the case in which the subsystems are uncoupled ($\strength = 0$), namely, $2/(3 \sqrt 3)$.
}
\label{fig:sync_window_vs_coupling}
\end{center}
\end{figure*}

If the coupling strength $\strength$ is increased, then, as Fig.~\ref{fig:sync_window_vs_coupling} illustrates, the S-shaped curves in Fig.~\ref{fig:master_slave_linear}(b) are stretched vertically, which increases the length of the synchronizing window. In fact, one can calculate the synchronizing window $S$ for System~\eqref{eq:master_slave_linear}; the result is $$S = \left [\frac{2 \left (1 - 3 \strength \right )}{3 \sqrt 3}, \frac{2\left (1 + 3 \strength \right )}{3 \sqrt 3}  \right ],$$ which is illustrated in Fig.~\ref{fig:sync_window_vs_coupling}(c). The implication of this expression is that strengthening the coupling makes synchronized regime shifts more likely, in the sense that more paths in parameter space lead to synchronized regime shifts. 

\section{Other simple couplings}

\begin{figure*}[htb]
\begin{center}
\includegraphics{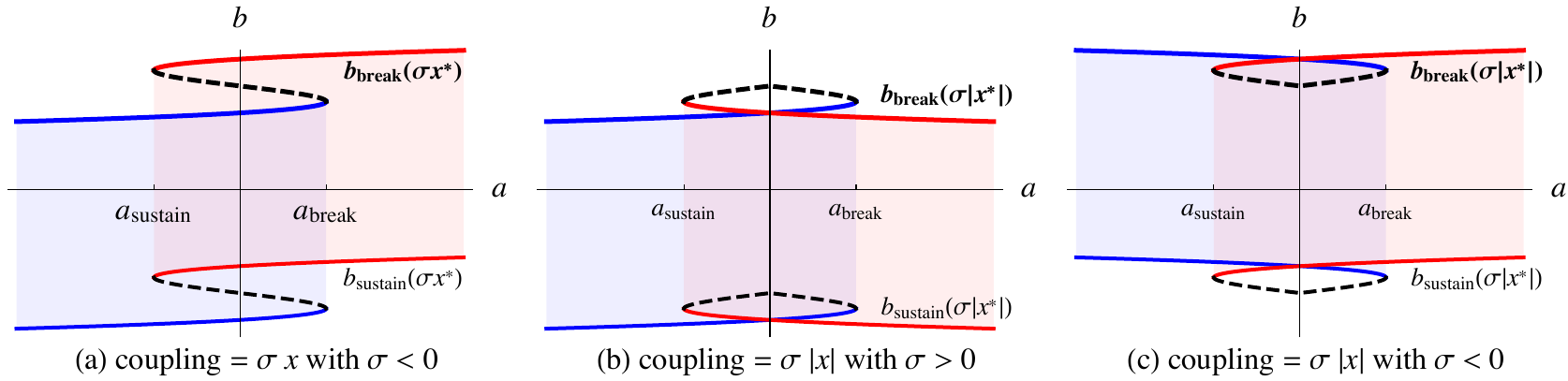}
\caption{{\bf Effect of other couplings on the slave subsystem's bifurcation diagram.} As in Fig.~\ref{fig:master_slave_linear}(b), if the master subsystem's equilibrium $x^*$ lies on its lower (respectively, upper) stable branch depicted in Fig.~\ref{fig:isolated}(b), then the saddle-node bifurcations of the slave subsystem, $b_\text{break}(\strength x^*)$ and $b_\text{sustain}(\strength x^*)$, are the blue (respectively, red) curves; the dashed curves correspond to the master subsystem being on its middle, unstable branch of equilibria. If there were no coupling, i.e., if $\sigma$ were equal to $0$, then the saddle-node bifurcations $b_\text{break}(\strength x^*)$ and $b_\text{sustain}(\strength x^*)$ would be given by the intersections of the black-dashed curves and the $a=0$ axis; comparing the blue and red curves with these intersections determines whether the master subsystem facilitates or inhibits a regime shift in the slave subsystem. 
Panel (a): The coupling \new{$\coupling{x}{y}{Y} = \strength x$} 
with $\strength = - 0.1$ makes it more difficult for the slave subsystem to cross its break point $b_\text{break}(\strength x^*)$ when the master subsystem has crossed its break point ($x^*>0$), and vice versa. Panel (b): The coupling \new{$\coupling{x}{y}{Y} = \strength |x|$} with $\strength = 0.1$ makes it easier for the slave subsystem to cross its break point $b_\text{break}(|\strength x^*|)$ no matter the sign of $x$. However, once the slave subsystem has crossed its break point, the value of $b$ must be reduced considerably more in order to cross the ``sustain'' saddle-node bifurcation, $b_\text{sustain}(|\strength x^*|)$, because of the coupling $\sigma |x|$ with $\sigma >0$. Flipping the sign of $\sigma$, as shown in panel (c), reverses these effects.}
\label{fig:other_couplings}
\end{center}
\end{figure*}

The results for \new{coupling functions $\coupling{x}{y}{Y} = \strength x$ and $\coupling{y}{x}{X} = 0$} 
with $\strength < 0$ can be obtained by reflecting the S-shaped curves $b_\text{break}(\strength x^*)$ and $b_\text{sustain}(\strength x^*)$ about the lines $b = b_\text{break}(0)$ and $b=b_\text{sustain}(0)$, respectively. In this case, the master subsystem facilitates (respectively, impedes) the slave subsystem's sudden shift to its upper branch of equilibria when $x^*$ is on its lower (respectively, upper) branch of equilibria. The effects of coupling functions \new{$\coupling{x}{y}{Y} = \strength | x |$} 
can be obtained similarly. They are illustrated in Figure~\ref{fig:other_couplings}.

\section{Facebook subgraph induced by all countries with protests}

\begin{figure*}[htbp]
\begin{center}
\includegraphics[width=\textwidth]{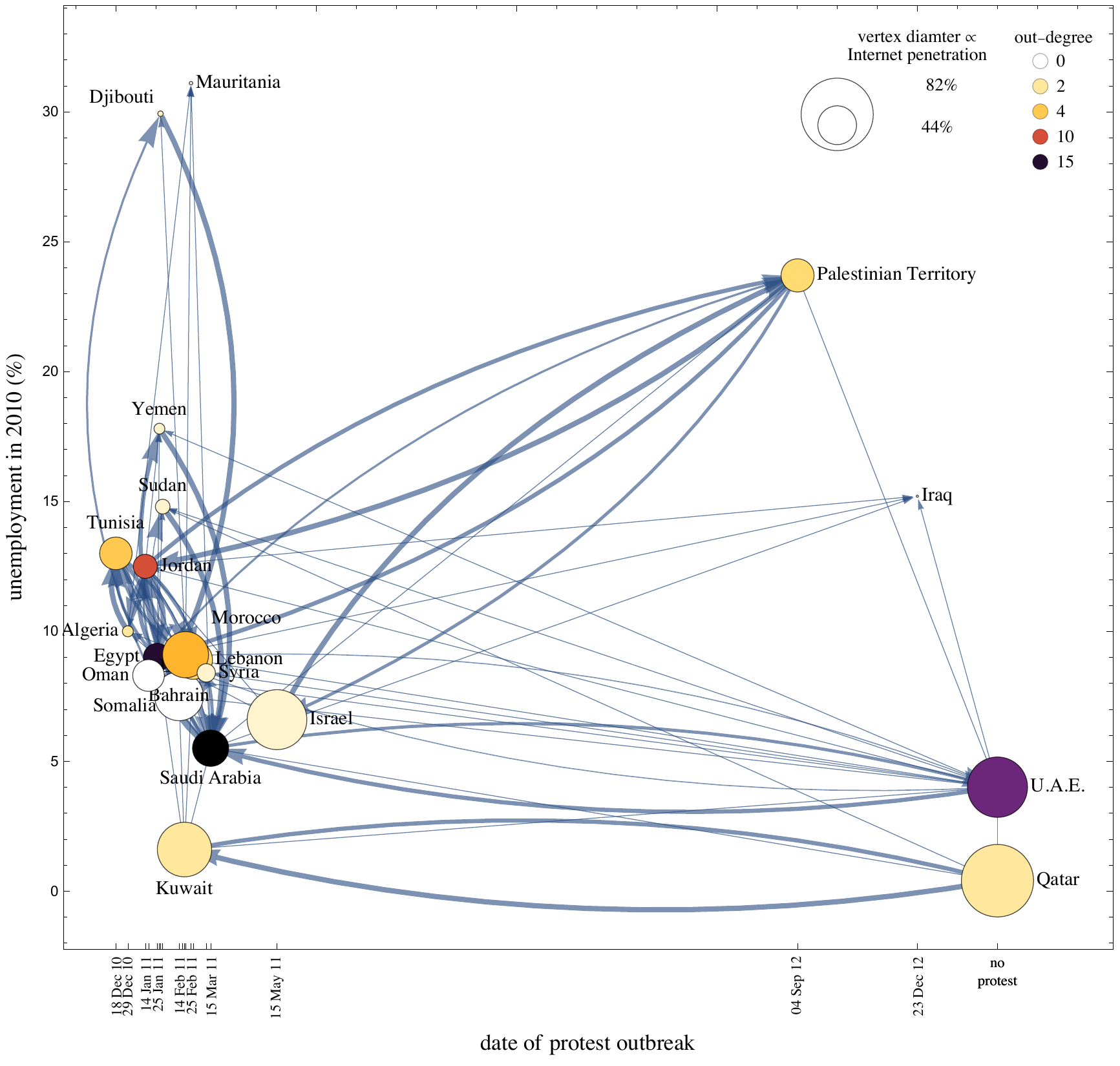}
\caption{
Facebook graph with countries excluded from Fig.~\ref{fig:Facebook} (namely, Djibouti, Israel, Palestinian Territory, Iraq, and Mauritania). The 2010 unemployment rate for Djibouti ($30\%$) was estimated from a linear regression between the available World Bank data~\cite{WorldBankData} and the ``fuzzy'' data in Table~1 of~\cite{Hussain:2013ir} (in which they estimate missing data by comparison with similar countries).}
\label{fig:Facebook_all}
\end{center}
\end{figure*}

In Fig.~\ref{fig:Facebook}, we excluded Djibouti because it does not have unemployment data, and we excluded Israel, Palestinian Territory, and Iraq because their protests began much later (May 2011, September 2012, and December 2012, respectively). We also excluded Mauritania because its unemployment was so large ($31.1\%$) that including it in Fig.~\ref{fig:Facebook} would obscure the rest of the data. 

Figure~\ref{fig:Facebook_all} replicates Fig.~\ref{fig:Facebook} with these countries excluded in Fig.~\ref{fig:Facebook}. Note that Djbouti and Mauritania have very small Internet penetration ($6.5\%$ and $4\%$, respectively), and they participate little in the Facebook subgraph shown in Fig.~\ref{fig:Facebook_all} (Djibouti has in-degree $2$ and out-degree $1$, while Mauritania has in-degree $3$ and out-degree $0$). Thus, we do not expect that excluding them from Fig.~\ref{fig:Facebook} has significant effects on the results in Sec.~\ref{sec:contagion_common_cause}. 

\section{Properties of countries in the hop motifs}
\label{sec:hop_motif_properties}

\begin{figure*}[htbp]
\begin{center}
\includegraphics[width=\textwidth]{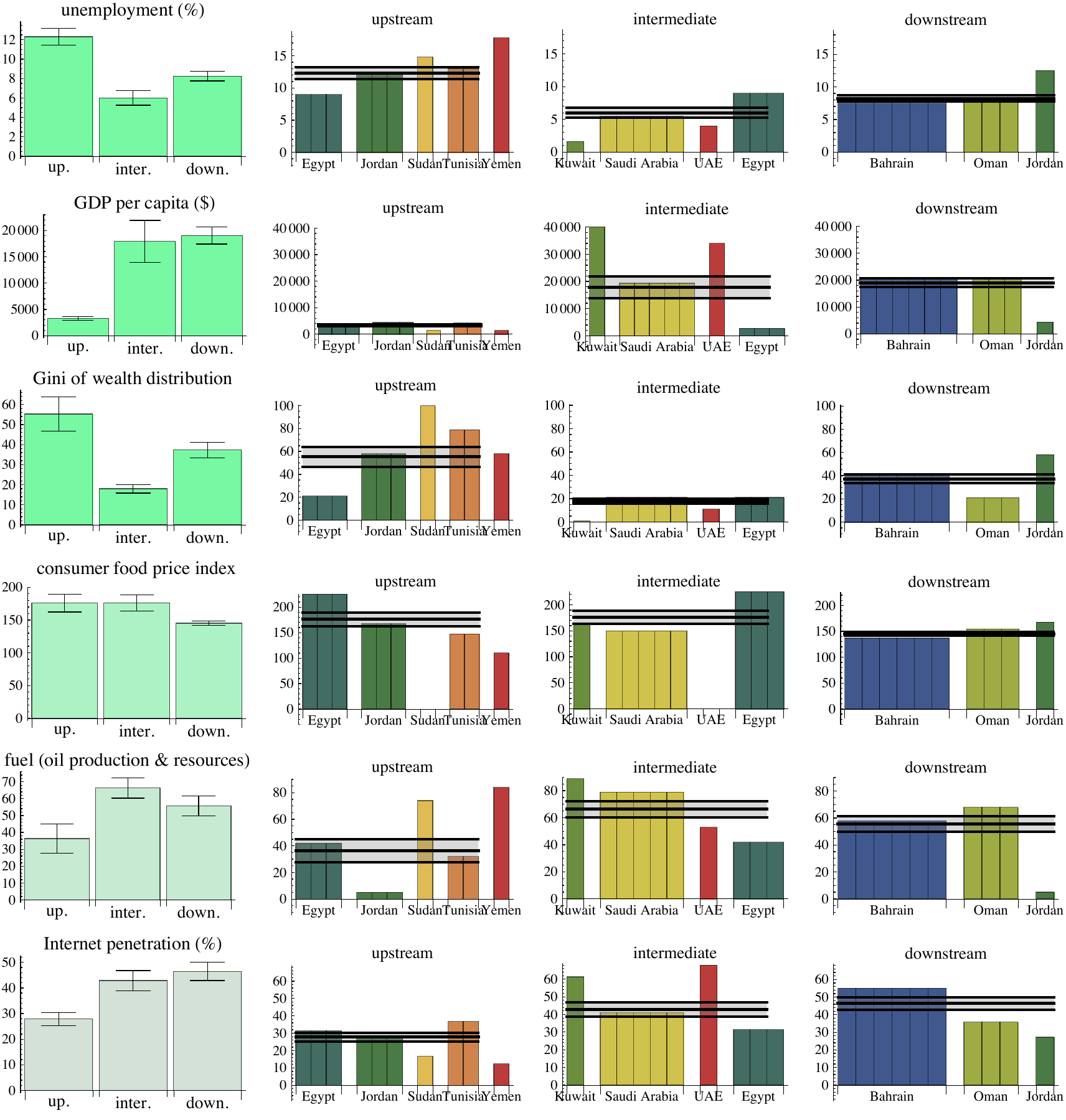}
\caption{{\bf Economic properties of countries in the ten hop motifs} (Table~\ref{tab:hopmotifs}). Each row is one attribute of a country (unemployment, etc.) The left-hand column are weighted averages of the countries in the different roles $\up, \inter, \down$ (labeled upstream, intermediate, downstream) in the hop motif (see Definition~\ref{def:hop}). The right-hand three columns show the countries in each of those roles; in each of these plots, the thick, horizontal line shows the mean, while the thin lines show the mean $\pm 1$ standard error. All data are from 2010.}
\label{fig:motif:economic}
\end{center}
\end{figure*}

\begin{figure*}[htbp]
\begin{center}
\includegraphics[width=\textwidth]{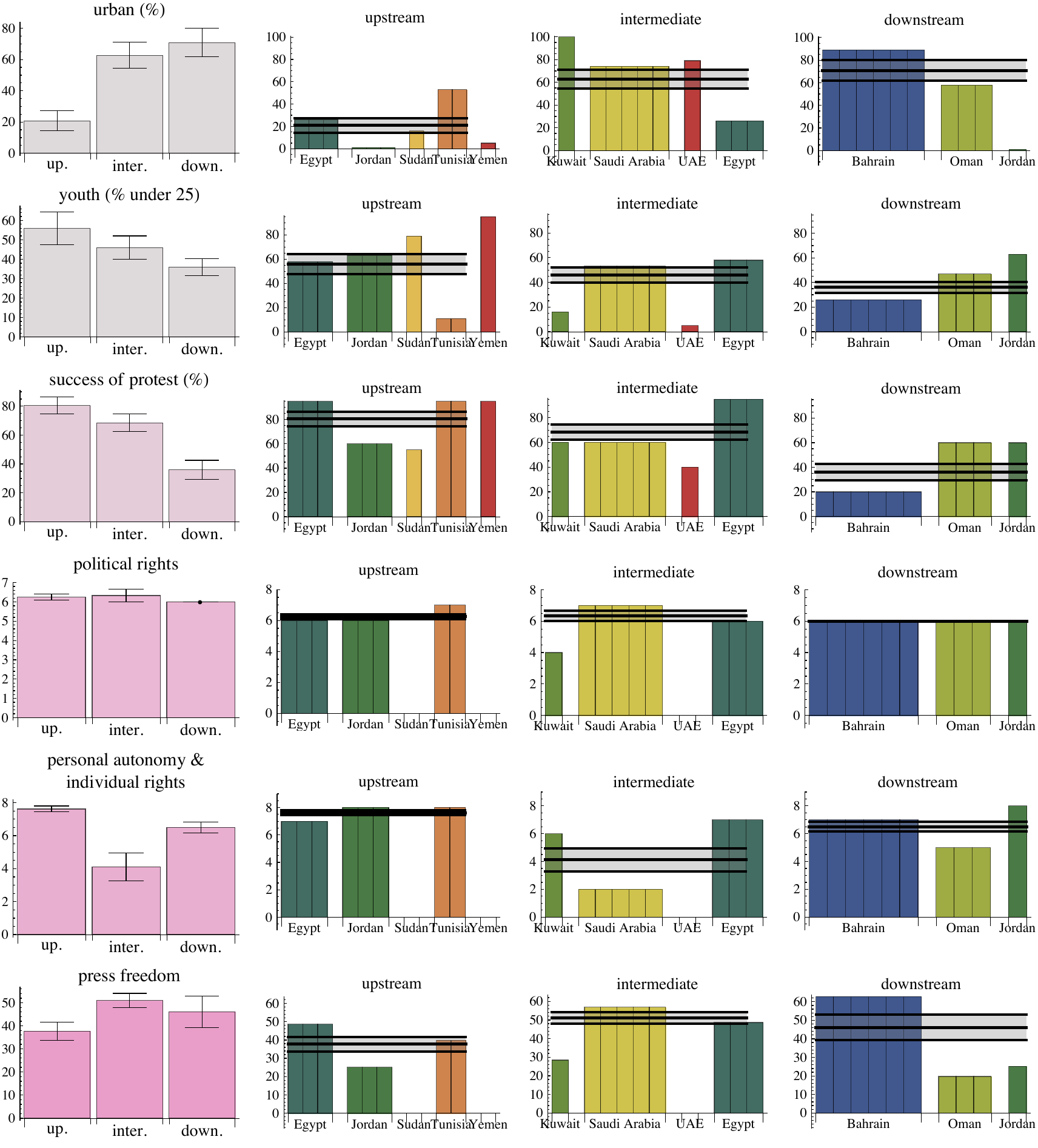}
\caption{
{\bf Social and political properties of countries in the ten hop motifs} (Table~\ref{tab:hopmotifs}). 
As in Fig.~\ref{fig:motif:economic}, the left-hand column shows weighted averages for the countries in the three roles $\up, \inter, \down$ in the hop motifs (as defined in Definition~\ref{def:hop}), while the other columns show the countries (and their properties) in those weighted averages. All the data are from 2010 except for press freedom (2013).
}
\label{fig:motif:social}
\end{center}
\end{figure*}

Figures~\ref{fig:motif:economic} and~\ref{fig:motif:social} show properties of the countries in these different roles $\up, \inter, \down$ in the hop motifs. 
The ``upstream'' countries $\up$ appear to be relatively close to their tipping points because of their relatively high unemployment, high economic inequality, low GDP per capita, and large youth bulges. Recall that intermediate countries $\inter$ may have spread influence to protest from these upstream countries $\up$ to ``downstream'' countries $\down$. Country $\down$ began to protest before $\inter$ did, perhaps because $\down$ was closer to its tipping point [e.g., downstream countries $\down$ had significantly higher unemployment and greater economic inequality (Gini coefficient)] and because intermediate countries tend to have relatively strong economies and significant revenue from oil, indicating a large distance from a tipping point. Intermediate and downstream countries had significantly higher Internet and mobile phone penetration, indicating their greater susceptibility to influence from ongoing protests.

Political rights and civil liberties~\cite{PoliticalFreedomData} are too coarse-grained to distinguish among these countries. However, some more specific measures, such as personal autonomy and individual rights, show greater variance among countries involved in the Arab Spring; the fifth row of Fig.~\ref{fig:motif:social} shows that intermediate countries $\inter$ (Kuwait, Saudi Arabia, Egypt; data for U.A.E. is missing) tend to have greater personal autonomy and individual rights, which may play a role in delaying their protests. However, even when these populations were not yet protesting, these citizens may nevertheless have been communicating inspiration to protest to other countries. Consistent with this interpretation, these intermediate countries also had greater freedom of the press (bottom row of Fig.~\ref{fig:motif:social}).

\end{document}